\documentclass[twocolumn, switch]{article}
\usepackage{arxiv}

\usepackage[utf8]{inputenc} 
\usepackage[T1]{fontenc}    
\usepackage{hyperref}       
\usepackage{url}            
\usepackage{booktabs}       
\usepackage{amsfonts}       
\usepackage{nicefrac}       
\usepackage{microtype}      
\usepackage{lipsum}		    
\usepackage{graphicx}
\usepackage[numbers,square,sort&compress]{natbib} 
\usepackage{authblk}
\usepackage{lipsum}			

\usepackage{subfig}
\usepackage{graphicx}
\usepackage{multirow}
\usepackage{booktabs} 
\usepackage{hyperref}
\usepackage{xspace}
\usepackage{amsmath}

\newcommand{\ie}{\emph{i.e.}\xspace}
\newcommand{\eg}{\emph{e.g.}\xspace}
\newcommand{\etal}{\emph{et al.}\xspace}

\definecolor{tmi_blue_color}{rgb}{0.086, 0.55, 0.84}
\definecolor{cadmiumgreen}{rgb}{0.0, 0.42, 0.24}

\hypersetup{
    colorlinks=true,
    linkcolor=blue,
    filecolor=blue,      
    urlcolor=blue,
    }

\makeatletter
\newcommand{\clearsubcaptcounter}{\setcounter{sub\@captype}{0}}
\makeatother

\usepackage{titlesec}
\titlespacing\section{0pt}{12pt plus 3pt minus 3pt}{1pt plus 1pt minus 1pt}
\titlespacing\subsection{0pt}{10pt plus 3pt minus 3pt}{1pt plus 1pt minus 1pt}
\titlespacing\subsubsection{0pt}{8pt plus 3pt minus 3pt}{1pt plus 1pt minus 1pt}

 \usepackage{tikz,xcolor,hyperref}

 \definecolor{lime}{HTML}{A6CE39}
 \DeclareRobustCommand{\orcidicon}{
 	\begin{tikzpicture}
 	\draw[lime, fill=lime] (0,0) 
 	circle [radius=0.16] 
 	node[white] {{\fontfamily{qag}\selectfont \tiny ID}};
 	\draw[white, fill=white] (-0.0625,0.095) 
 	circle [radius=0.007];
 	\end{tikzpicture}
 	\hspace{-2mm}
 }
 \foreach \x in {A, ..., Z}{\expandafter\xdef\csname orcid\x\endcsname{\noexpand\href{https://orcid.org/\csname orcidauthor\x\endcsname}
 			{\noexpand\orcidicon}}
 }

\title{Convolutional Neural Network to Restore Low-Dose Digital Breast Tomosynthesis Projections in a Variance Stabilization Domain}

 \newcommand{\shorttitle}{CNN to restore low-dose DBT projections in a variance stabilization domain}

\author[1,2]{Rodrigo B. Vimieiro}
\author[2]{Chuang Niu}
\author[3,4]{Hongming Shan}
\author[5]{Lucas R. Borges}
\author[2,*]{Ge Wang}
\author[1,*]{Marcelo A. C. Vieira}

\affil[1]{Department of  Electrical and Computer Engineering, S\~{a}o Carlos School of Engineering, University of S\~{a}o Paulo, S\~{a}o Carlos, Brazil}
\affil[2]{Department of Biomedical Engineering, School of Engineering, Center
for Biotechnology and Interdisciplinary Studies, Rensselaer Polytechnic
Institute, Troy, NY, USA}

\affil[3]{Institute of Science and Technology for Brain-inspired Intelligence, Fudan University, Shanghai, China}

\affil[4]{Shanghai Center for Brain Science and Brain-inspired
Technology, Shanghai, China}

\affil[5]{Department of Medical Imaging, Hematology and Clinical Oncology, Ribeir\~{a}o Preto School of Medicine, University of S\~{a}o Paulo, Ribeir\~{a}o Preto, Brazil}

\affil[*]{Co-corresponding authors}

\hypersetup{
pdftitle={},
pdfsubject={},
pdfauthor={},
pdfkeywords={},
}

\begin{document}

\twocolumn[ 
  \begin{@twocolumnfalse} 
  
\maketitle


\begin{abstract}

Digital breast tomosynthesis (DBT) exams should utilize the lowest possible radiation dose while maintaining sufficiently good image quality for accurate medical diagnosis. 
In this work, we propose a convolution neural network (CNN) to restore low-dose (LD) DBT projections to achieve an image quality equivalent to a standard full-dose (FD) acquisition. The proposed network architecture benefits from priors in terms of layers that were inspired by traditional model-based (MB) restoration methods, considering a model-based deep learning approach, where the network is trained to operate in the variance stabilization transformation (VST) domain. To accurately control the network operation point, in terms of noise and blur of the restored image, we propose a loss function that minimizes the bias and matches residual noise between the input and the output. The training dataset was composed of clinical data acquired at the standard FD and low-dose pairs obtained by the injection of quantum noise. The network was tested using real DBT projections acquired with a physical anthropomorphic breast phantom. The proposed network achieved superior results in terms of the mean normalized squared error (MNSE), training time and noise spatial correlation compared with networks trained with traditional data-driven methods. The proposed approach can be extended for other medical imaging application that requires LD acquisitions. 

\end{abstract}


\vspace{0.35cm}

  \end{@twocolumnfalse} 
] 

\section{Introduction}
\label{sec:introduction}

\begin{figure*}[ht]
    \centering
    \includegraphics[width=1\linewidth]{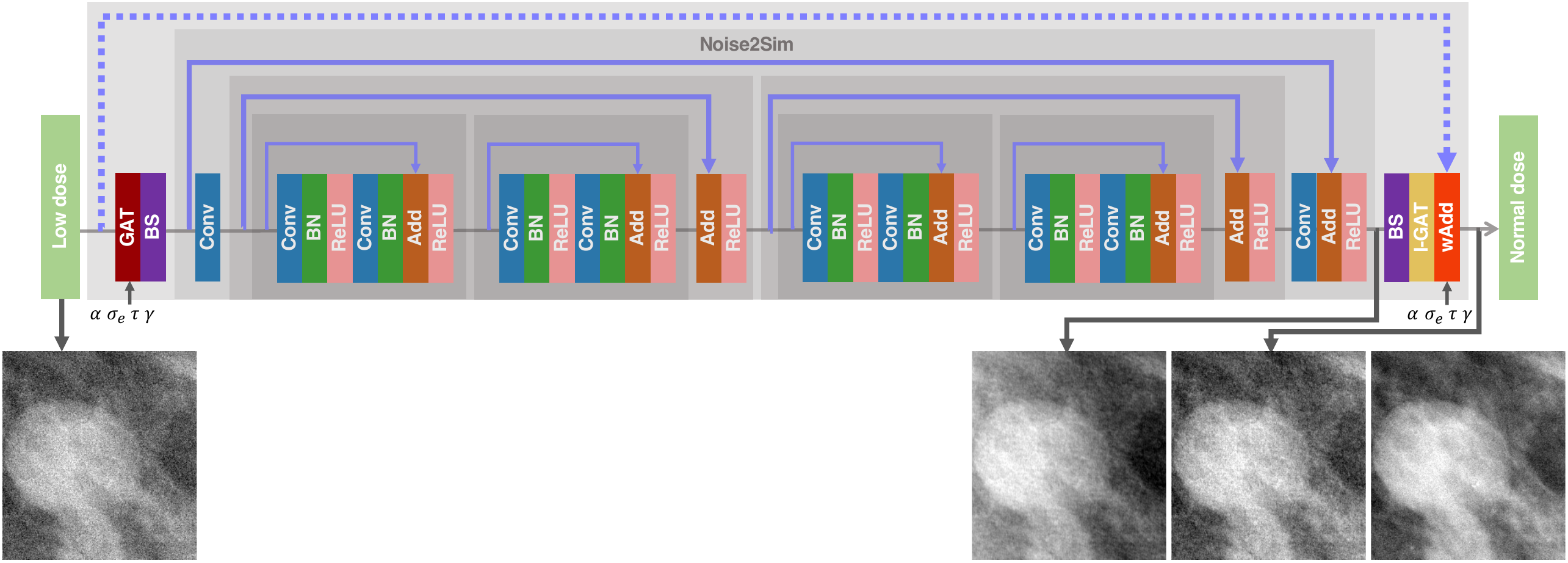}
    \caption{Illustration of the proposed network architecture. From left to right, ROIs represent the low-dose image, estimated noise-free image, restored image and the respective full-dose.
    }
    \label{fig:network}
\end{figure*}

Image restoration has been an active topic, especially in the field of medical imaging systems where radiation dose is a concern. Digital breast tomosynthesis (DBT) is an imaging modality currently being used for breast screening. In this system,  the x-ray tube moves along an arc acquiring several projections of the breast. In the end, a pseudo-3D volume is reconstructed by a combination of the set of projections, and radiologists may search for signs of breast cancer~\cite{lai2020digital}.

Since this technique is used on a population-based breast cancer screening procedure and the image acquisition process utilizes ionizing radiation, it is desirable to achieve low radiation doses and preserve image quality for satisfactory diagnosis. Restoring images from low-dose acquisitions has been an active research topic and has been solved by different approaches, such as traditional iterative reconstruction methods~\cite{das2010penalized, xu2015statistical, zheng2017detector} and denoising techniques~\cite{wu2012dose, borges2017pipeline, borges2018restoration}. More recently, data-based methods using neural networks have also been applied to the field, achieving promising results~\cite{wu2017iterative, kang2017deep, chen2017low1, wolterink2017generative, chen2017low2, kang2018deep, shan20183,  shan2019competitive, yin2019domain, wu2021drone}. 

Specifically, in the DBT area, Gao, Fessler and Chan~\cite{gao2021deep} proposed a Wasserstein generative adversarial network (WGAN) for denoising DBT slices. The network was trained on digital and physical phantoms, tested on clinical images and both mean squared error (MSE) and adversarial loss were used in the training step. Liu \etal~\cite{liu2018radiation} used two breast specimens to train a convolutional neural network (CNN) model for the restoration of DBT images. The low-dose (LD) and full-dose (FD) image pairs were acquired in the equipment and the network was applied in the projections. Sahu \etal~\cite{sahu2019using} trained a GAN on synthetic images, generated through virtual clinical trials (VCT) software, to denoise LD DBT projections. Both LD and HD images were generated in the VCT software. 

All these works rely on the fact that the network is expected to learn all the restoration process strictly from the provided data. However, adding prior knowledge on the network architecture might improve network performance. This approach is known as model-based deep learning and has been investigated in previous works~\cite{wu2017iterative, kang2018deep, chen2018learn, wurfl2018deep, adler2018learned, gong2018pet, zhang2019vst, wu2021drone}. 

In this work, inspired by a model-based (MB) denoising pipeline~\cite{borges2018restoration}, designed to restore LD DBT projections, we propose a deep CNN architecture, that benefits from prior knowledge, to restore LD DBT projections. 

First, we pre-trained the network in a variance-stabilizing transformation (VST) domain with a self-learning framework~\cite{niu2020noise2sim}. 

In common training strategies, local patches are extracted from the image and feeded in the network. This makes it hard for the network to know the spatial noise dependencies existing in the mammography image. In the VST domain, the denoising task becomes easier as the transformation is intended to minimize the signal and spatial dependencies. In other words, this will facilitate the network learning process, such that it does not have to learn different denoising strengths, depending on the input.

As we do not have noise-free clinical images to use as references in the VST domain, we used the self-learning method to estimate this image. This allows the use of clinical images in the training phase by the network in the VST domain. 

In the second step, we loaded the pre-trained weights in an architecture that contains both the forward VST and its inverse as layers. Also, we added a weighted sum module in the last residual layer, so the restored image matches the expected mean and variance of the FD. 

Zhang \etal~\cite{zhang2019vst} proposed VST-Net to denoise natural images corrupted by Poisson noise. Their network performs the Anscombe transformation~\cite{anscombe1948transformation} and its inverse as convolutional operations. Our work differs by the facts that we used the generalized Anscombe transformation (GAT) due to the Poisson-Gaussian noise nature of DBT projections, we used a properly unbiased inverse and our transformations does not have learnable weights. Also, our last residual layer induces the network to match the mean and variance of the target image. 
We also proposed a loss function that yields a restored image residual noise (RN) similar to the target FD, with minimal bias. Narage \etal~\cite{nagare2021bias} proposed a similar loss function. Our approach differs in the fact that we directly calculate bias and RN from different image realizations. Also, we restricted this loss to fine-tune the final network so it can achieve specific denoising properties.  

The manuscript is divided as follows. Section~\ref{sec:Theoretical_Background} gives a general overview of the image formation, the transformations and the respective network. Section~\ref{sec:Materials_Methods} describe how the networks were trained, the loss functions used and the datasets. Section~\ref{sec:Results_Discussions} presents and discusses the results and finally, Section~\ref{sec:Conclusion} concludes the work, states the limitations on the work and shows future works. 

\section{Theoretical Background}
\label{sec:Theoretical_Background}

Noise in DBT projections is typically described by a Poisson-Gaussian model. Quantum noise accounts for the signal-dependent part, whereas electronic or thermal noise is described by the signal-independent part. The raw image acquisition at the standard full-dose (FD) can be formulated as:

\begin{equation}
    z(x) = \alpha(x) p(x) + n(x) + \tau
    \label{eq:FDNoiseModel}
\end{equation}

\noindent where $x$ indicates the pixel spatial coordinates, $p(x) \sim \mathcal{P}(\alpha(x)^{-1} y(x))$ is a random variable with Poisson distribution, $\alpha(x)$ is the spatially dependent quantum gain, $y(x)$ is the (unknown) noise-free image, $n(x) \sim \mathcal{N}(0, \sigma^2_e)$ is a random variable with Gaussian distribution, $\sigma^2_e$ is the variance of the electronic noise and $\tau$ is the pixel offset. 

One can model a lower-dose (LD) projection $z_\gamma(x)$ by fixing the radiographic factors as in $z(x)$ except for a reduction in the current-time product (mAs):

\begin{equation}
    z_\gamma (x) = \alpha(x) p_\gamma(x) + n(x) + \tau
    \label{eq:LDNoiseModel}
\end{equation}

\noindent where $p_\gamma \sim \mathcal{P}(\gamma \alpha(x)^{-1} y(x))$ and $0<\gamma<1$ is the corresponding dose reduction factor.

Such dose reduction directly impacts on the SNR, such as, for $z_i$:

\begin{equation}
    SNR = \frac{E\{z_i \,|\, y_i\}^2}{var\{z_i \,|\, y_i\}} = \frac{y_i^2}{\alpha_i^{-1} y_i + \sigma^2_e},
    \label{eq:FD_SNR}
\end{equation}

\noindent and if we subtract the pixel offset and divide it by $\gamma$, it is clear that the SNR is lower for $[z_\gamma]_i$:

\begin{equation}
    SNR_\gamma = \frac{E\{[z_\gamma]_i \,|\, \gamma y_i\}^2}{var\{[z_\gamma]_i \,|\, \gamma y_i\}} = \frac{y_i^2}{\gamma^{-1} \alpha_i^{-1} y_i + \sigma^2_e \gamma^{-2}}.
    \label{eq:LD_SNR}
\end{equation}

In Borges \etal~\cite{borges2018restoration}, an image restoration pipeline was proposed to restore the LD DBT projections and achieve the FD SNR.

The restoration process can be treated by a modular approach, where first, a VST is applied to approximate noise to a unitary variance. In the case of  Poisson-Gaussian distributions, the GAT~\cite{starck1998image} is commonly used:

\begin{equation}
f(\dot{z}_i) = 
\begin{cases}
2 \sqrt{\dot{z}_i + \frac{3}{8} + \sigma_e^2}, & \text{ if } \dot{z}_i>-\frac{3}{8}-\sigma_e^2  \\ 
0 & \text{ if } \dot{z}_i\leq -\frac{3}{8}-\sigma_e^2,
\end{cases}
\label{eq:GAT}
\end{equation}

\noindent where $\dot{z}_i$ is the observed pixel after an affine transformation which subtracts the offset and divides the quantum noise gain and  $\dot{\sigma}_e^2$ is the electronic noise variance after the affine transformation.  

As the GAT is a non-linear transformation, removing noise on its domain may introduce bias when using the algebraic inverse~\cite{makitalo2012optimal}. This one can be solved by applying the asymptotically unbiased inverse, however,~\cite{makitalo2012optimal} demonstrated that such approach is not sufficient for low-intensity images, and they proposed a closed-form approximation of the exact unbiased inverse: 

\begin{equation}
\widetilde{\mathcal{I}}(d) = \frac{1}{4} d^2 + \frac{1}{4} \sqrt{\frac{3}{2}} d^{-1} - \frac{11}{8}  d^{-2} + \frac{5}{8} \sqrt{\frac{3}{2}} d^{-3} - \frac{1}{8} - \sigma_e^2,
\label{eq:iGAT}
\end{equation}
 
 \noindent where $d$ is the denoised image. Note that the affine inverse transformation is applied after~\eqref{eq:iGAT}.   
 
 As our primary goal is to match the signal and noise properties of LD and FD image, it is desired that their expected mean and variance match. So, after denoising the image in the VST domain, we have an estimate of the noise-free signal $[\widehat{y_\gamma}]_i$, then a weighted sum is performed between the input image and the denoised one:
 
 \begin{equation}
 \hat{z} = w_i \times ([z_\gamma]_i - \tau) + \bar{w_i} \times ([\widehat{y_\gamma}]_i - \tau) + \tau,
 \label{eq:wSum1}
 \end{equation}
 
 \noindent where
 
 \begin{equation}
     w_i = \sqrt{\frac{\alpha_i y_i  + \sigma_e}{\gamma \alpha_i y_i  + \sigma_e}},
 \label{eq:wSum2}
 \end{equation}
 
 \noindent and 
 
 \begin{equation}
     \bar{w_i} = \frac{1}{\alpha_i} - w_i,
 \end{equation}
 
 \noindent is sufficient to make the restored image expected mean and variance match the FD, considering the denoising was successfully~\cite{borges2018restoration}. As we do not have access to $y$, we approximate it to:
 
 \begin{equation}
 \hat{y} = \frac{[\widehat{y_\gamma}]_i - \tau}{\gamma}
 \end{equation}

Figure~\ref{fig:network} illustrates the proposed network architecture, where GAT block implements~\eqref{eq:GAT} within the affine transformation, BS normalize the signal from 0 to 1, iGAT implements~\eqref{eq:iGAT} and wAdd implements~\eqref{eq:wSum1} and~\eqref{eq:wSum2} with a residual layer. Note that, in clinical images, as we do not have access to $y_i$ from~\eqref{eq:FDNoiseModel}, we used the self-learning framework Noise2Sim~\cite{niu2020noise2sim} to estimate it and pre-train the network. 

\section{Materials \& Methods}
\label{sec:Materials_Methods}

\subsection{CNN Training}

In general, two approaches were investigated in this work. The first one is related to training the CNN to restore the LD image to the corresponding HD with the slightly modified Residual Network (ResNet) presented before~\cite{shan2021lossfunction}. We call it ResNet throughout the paper for the sake of simplicity. The second approach involves training in two phases, where it first learns how to remove all the noise from the image and, later, it is fine-tuned to closely match the noisy properties to its corresponding FD.  

\subsubsection{Frameworks}

We call the first framework the data-based learning (DBL) approach. Overall, three networks were trained with different datasets to investigate their performance. The first network was previously presented at~\cite{shan2021lossfunction}, where it was trained on FFDM images and at this work tested in the DBT dataset. For the second approach, we used the same network as in the first case, however, fine-tuning was performed in the DBT training dataset. The third network was trained and tested only on DBT data. In all these approaches, no previous information about the data was provided to the network, so it has to learn the restoration only from the provided data. 

The second framework provides previous knowledge for the network, so we call it hybrid learning (HBL). In this domain, we investigated two different methods. The first one, the network learns how to estimate the noise-free signal inside the VST domain. In this task, we used the Noise2Sim (N2S) framework~\cite{niu2020noise2sim}, which uses non-local image information to estimate a noise-free image in an unsupervised manner. Second, we used the previous network as a pre-trained model and fine-tuned it with a bias-residual-noise (BRN) loss function after the whole pipeline.

\subsubsection{Loss Functions}

The importance of the loss function on image restoration was previously presented at~\cite{zhao2016loss} for natural images and at~\cite{shan2021lossfunction} for medical images. In this work, we used two commonly known mean absolute error (MAE) and perceptual loss (PL) functions. Also, we propose the bias-residual-noise loss function which was inspired by the decomposition on the MNSE proposed at~\cite{borges2018restoration}. 

The BRN loss function tries to match the RN between the pair of LD and FD images while it minimizes the signal bias of the restored image. \eqref{eq:BRNloss} shows the composition of these factors in the loss function:

\begin{equation}
    \mathcal{L}_{\mathcal{B}\mathcal{R}_{\mathcal{N}}} = \mathcal{B}^2 + \lambda_{\mathcal{R}_{\mathcal{N}}} *  |\mathcal{R}_{\mathcal{N}_{FD}} - \mathcal{R}_{\mathcal{N}_{RD}}|,
    \label{eq:BRNloss}
\end{equation}

\noindent where $\lambda_{\mathcal{R}_{\mathcal{N}}}$ is a weighting factor to control the denoising strength by imposing a match between the RN of the FD and the restored dose (RD). 

\subsection{Dataset}

\begin{figure*}[ht]
	\centering	
	\subfloat{\includegraphics[scale=2.3]{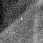}}
	\enspace
	\subfloat{\includegraphics[scale=2.3]{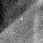}}
	\enspace
	\subfloat{\includegraphics[scale=2.3]{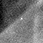}}
	\enspace
	\subfloat{\includegraphics[scale=2.3]{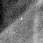}}
	\enspace
	\subfloat{\includegraphics[scale=2.3]{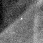}}
	\enspace
	\subfloat{\includegraphics[scale=2.3]{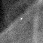}}
	
	\subfloat {\includegraphics[scale=2.3]{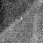}}
	\enspace
	\subfloat {\includegraphics[scale=2.3]{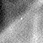}}
	\enspace
	\subfloat {\includegraphics[scale=2.3]{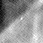}}
	\enspace
	\subfloat {\includegraphics[scale=2.3]{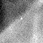}}
	\enspace
	\subfloat {\includegraphics[scale=2.3]{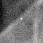}}
	\enspace
	\subfloat{\includegraphics[scale=2.3]{imgs/rois/phantom/60_01_Mammo_R_CC.png}}
	
	\subfloat {\includegraphics[scale=2.3]{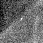}}
	\enspace
	\subfloat {\includegraphics[scale=2.3]{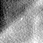}}
	\enspace
	\subfloat {\includegraphics[scale=2.3]{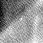}}
	\enspace
	\subfloat {\includegraphics[scale=2.3]{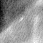}}
	\enspace
	\subfloat {\includegraphics[scale=2.3]{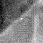}}
	\enspace
	\subfloat {\includegraphics[scale=2.3]{imgs/rois/phantom/60_01_Mammo_R_CC.png}}
	
	\clearsubcaptcounter
	\subfloat[LD] {\includegraphics[scale=2.3]{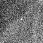}}
	\enspace
	\subfloat[FFDM] {\includegraphics[scale=2.3]{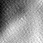}}
	\enspace
	\subfloat[FFDM$^*$] {\includegraphics[scale=2.3]{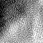}}
	\enspace
	\subfloat[DBT] {\includegraphics[scale=2.3]{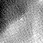}}
	\enspace
	\subfloat[MB] {\includegraphics[scale=2.3]{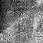}}
	\enspace
	\subfloat[FD] {\includegraphics[scale=2.3]{imgs/rois/phantom/60_01_Mammo_R_CC.png}}
	
	\caption{Visualization of an ROI for a visual comparison of the DBL restorations. Where (a) represents the low-dose image, (b) FFDM network restoration, (c) FFDM with transfer-learning, (d) DBT network, (e) the MB method and (f) the corresponding full-dose image.}
	\label{fig:DBL_roi_proj_phantom}
\end{figure*}

We used different datasets to train and test DBL and HBL approaches. 

\subsubsection{Training}

\paragraph{DBL} In this approach, we used two datasets. As, in this work, we have a reduction factor as low as 5\%, we had to train FFDM networks for these new values. For that, we used the same FFDM dataset used in~\cite{shan2021lossfunction} for training. It consists of 400 retrospective clinical images acquired at the Barretos Cancer Hospital (Brazil) under IRB approval (\textit{Certificado de Apresenta\c{c}\~{a}o para Aprecia\c{c}\~{a}o \'Etica} - CAAE \#78625417.1.1001.5437). For the DBT network, we used a new dataset, consists of 1,982 retrospective clinical exams (29,730 projections) acquired at the Institute of Radiology (InRad), Faculty of Medicine, University of S\~{a}o Paulo (Brazil) under IRB approval (CAAE \#56699016.7.0000.0065). Both datasets were generated using the Hologic Selenia Dimensions equipment (Hologic, Bedford, MA) and the images were used as raw data. All clinical images were carefully anonymized to preserve patients' medical records.

For FFDM and DBT images, we simulated LD acquisitions using the method proposed at~\cite{borges2016method, borges2017method}, which injects quantum and electronic noise in the VST domain. 

\paragraph{HBL} In this procedure, we used the aforementioned DBT dataset for the N2S framework and we also generated virtual clinical images using the virtual clinical trial (VCT) framework developed at the University of Pennsylvania~\cite{BakicVCT}. We used that framework so that we could generate several acquisitions of the same patient with different noise realizations. 7 patients were simulated and we generated the corresponding noise-free DBT projections. Note that it is possible to simulate as many patients as desired. We simulated five noise realizations for each projection set using the method proposed at~\cite{borges2019noise}. The geometry of the Hologic Selenia Dimensions system was used and noise simulation accounts for the quantum and electronic noise and also a correlation kernel. 

\subsubsection{Testing}

To test all the networks in a real scenario, we used a dataset of an anthropomorphic phantom that was designed by the University of Pennsylvania~\cite{carton2011development} and produced by CIRS, Inc. (Reston, VA). We performed 60 DBT acquisitions that produced 20 images at the FD and 10 images at 50\%, 25\%, 15\% and 5\% of the FD. The standard radiographic factors were obtained by using the automatic exposure control (AEC), yielding 31 kVp and 60 mAs. In the system manual mode, by reducing accordingly the mAs, it is possible to achieve LD images. This dataset was used in both DBL and HBL approaches. 

\subsection{Implementation Details}

\begin{figure*}[ht]
	\centering	
	\subfloat{\includegraphics[scale=2.3]{imgs/rois/phantom/30_01_Mammo_R_CC.png}}
	\hfill
	\subfloat{\includegraphics[scale=2.3]{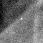}}
	\hfill
	\subfloat{\includegraphics[scale=2.3]{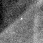}}
	\hfill
	\subfloat{\includegraphics[scale=2.3]{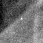}}
	\hfill
	\subfloat{\includegraphics[scale=2.3]{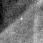}}
	\hfill
	\subfloat{\includegraphics[scale=2.3]{imgs/rois/phantom/MB_30_01_Mammo_R_CC.png}}
	\hfill
	\subfloat{\includegraphics[scale=2.3]{imgs/rois/phantom/60_01_Mammo_R_CC.png}}
	
	\subfloat {\includegraphics[scale=2.3]{imgs/rois/phantom/15_01_Mammo_R_CC.png}}
	\hfill
	\subfloat {\includegraphics[scale=2.3]{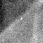}}
	\hfill
	\subfloat {\includegraphics[scale=2.3]{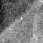}}
	\hfill
	\subfloat {\includegraphics[scale=2.3]{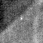}}
	\hfill
	\subfloat {\includegraphics[scale=2.3]{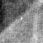}}
	\hfill
	\subfloat {\includegraphics[scale=2.3]{imgs/rois/phantom/MB_15_01_Mammo_R_CC.png}}
	\hfill
	\subfloat{\includegraphics[scale=2.3]{imgs/rois/phantom/60_01_Mammo_R_CC.png}}

	\subfloat {\includegraphics[scale=2.3]{imgs/rois/phantom/9_01_Mammo_R_CC.png}}
	\hfill
	\subfloat {\includegraphics[scale=2.3]{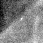}}
	\hfill
	\subfloat {\includegraphics[scale=2.3]{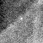}}
	\hfill
	\subfloat {\includegraphics[scale=2.3]{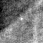}}
	\hfill
	\subfloat {\includegraphics[scale=2.3]{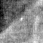}}
	\hfill
	\subfloat {\includegraphics[scale=2.3]{imgs/rois/phantom/MB_9_01_Mammo_R_CC.png}}
	\hfill
	\subfloat {\includegraphics[scale=2.3]{imgs/rois/phantom/60_01_Mammo_R_CC.png}}
	
	\clearsubcaptcounter
	\subfloat[LD] {\includegraphics[scale=2.3]{imgs/rois/phantom/3_01_Mammo_R_CC.png}}
	\hfill
	\subfloat[N2S] {\includegraphics[scale=2.3]{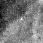}}
	\hfill
	\subfloat[VST-0.01] {\includegraphics[scale=2.3]{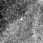}}
	\hfill
	\subfloat[VST-0.34] {\includegraphics[scale=2.3]{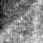}}
	\hfill
	\subfloat[VST-0.95]{\includegraphics[scale=2.3]{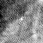}}
	\hfill
	\subfloat[MB]{\includegraphics[scale=2.3]{imgs/rois/phantom/MB_3_01_Mammo_R_CC.png}}
	\hfill
	\subfloat[FD]{\includegraphics[scale=2.3]{imgs/rois/phantom/60_01_Mammo_R_CC.png}}

	\caption{Visualization of an ROI for a visual comparison of the HBL restorations. Where (a) represents the low-dose image, (b) N2S network restoration, (c) to (e) VST networks with its respective $\lambda_{RN}$ values, (f) the MB method and (f) the corresponding full-dose image.}
	\label{fig:HBL_roi_proj_phantom}
\end{figure*}

All neural networks were implemented using PyTorch Deep Learning library~\footnote{\url{www.pytorch.org}}, trained and tested using an NVIDIA TITAN Xp with 12GB of RAM. However, specific details are related to each restoration framework.

\paragraph{DBL} In this framework approach, one neural network is trained for each reduction factor. We trained FFDM networks as done in~\cite{shan2021lossfunction}, \ie, first training a network with MAE loss with 60 epochs and LR equals $1\times10^{-4}$ and subsequently training a network with PL4 for 60 epochs and the LR reduced by ten times. The batch size was set to 256 in the pre-trained network and 64 in the later training. The LR was reduced by half every 10 epochs. For fine-tuning on DBT data, we used the aforementioned model as a pre-trained network and trained for 10 epochs, PL4 and LR equal to $5\times10^{-6}$. In this approach, the LR was reduced by half every two epochs. DBT network was trained in the same manner as the FFDM one, except for the fact that only DBT data was used. We used Adam optimizer with running averages of gradient and its square equal $0.5$ and $0.999$, respectively. All the networks were trained with 256,000 patches.

\paragraph{HBL} Different from the DBL, we restricted the network training procedure only for the 50\% of dose reduction. We observed that training at lower doses forces the network to have strong denoising properties and yield to a bias increasing. This model was tested on the images at the other reduction factors. In the N2S model, we trained the ResNet in an unsupervised way. The target images are generated by the N2S framework with a technique that was inspired by the non-local means method. The N2S training was performed in the VST domain, \ie, we performed the GAT on the data before training. We trained for 70 epochs, batch size of 128, Adam optimizer with the same parameters as in the DBL, LR which starts from 0, go up to $1\times10^{-4}$ and return to 0 (default method in the framework), MSE loss function and the number of simulated images equal to 8. 

After properly training the ResNet in the N2S framework, we loaded its weights in the VST network and performed a fine-tuning with the BRN loss function. 90,650 patches of the VCT projections were used in the training process. To correctly measure the BRN loss, we stacked the five noise realizations in sequence and set the batch size to 60. The loss was then calculated between the LD image and the corresponding FD, outside of the VST domain. We fine-tuned the model only for 2 epochs and set the LR to $1\times10^{-5}$. The $\lambda_{RN}$ was optimized and the results are presented in the following sections. 

\paragraph{MB} We also used as a benchmark the MB restoration technique that was presented in~\cite{borges2018restoration}. This approach implements that pipeline mentioned in Section~\ref{sec:Theoretical_Background} and uses the block-matching and 3D filtering (BM3D)~\cite{dabov2007image} in the VST domain.

\subsection{Figures of Merit}

\begin{table*}[ht]
\centering
\caption{MNSE analysis with its decomposition on bias and RN for the DBL approach and comparison with LD, FD and MB results. Values highlighted in blue mean the lowest bias for the DL restorations. In green, RN values which are closest to the value from FD images.}
\label{tab:MNSEdecomDBL}
\begin{tabular}{clccccccc} 
\toprule
mAs                 & Metric                      & Noisy    & DL-FFDM  & DL-FFDM* & DL-DBT  & MB      & \multicolumn{2}{l}{Full-dose}                                            \\ 
\hline
\multirow{3}{*}{30} & $\mathcal{B}^2$             & 0.15\%   & \textcolor{tmi_blue_color}{\textbf{0.50}}\%  & 0.62\%  & 0.87\%  & 0.26\%  & \multirow{4}{*}{$\mathcal{B}^2$}             & \multirow{4}{*}{0.09\%}   \\ 
\cline{2-7}
                    & $\mathcal{R}_{\mathcal{N}}$ & 24.86\%  & \textcolor{cadmiumgreen}{\textbf{12.37}}\%  & 10.12\% & 11.10\% & 13.23\% &                                              &                           \\ 
\cline{2-7}
                    & MNSE                        & 25.02\%  & 12.88\%  & 10.74\%  & 11.97\% & 13.50\% &                                              &                           \\ 
\cline{1-7}
\multirow{3}{*}{15} & $\mathcal{B}^2$             & 0.45\%  & 3.25\%   & 2.95\%   & \textcolor{tmi_blue_color}{\textbf{2.87}}\%  & 0.82\%  &                                              &                           \\ 
\cline{2-9}
                    & $\mathcal{R}_{\mathcal{N}}$ & 50.80\% & 8.47\%  & \textcolor{cadmiumgreen}{\textbf{8.73}}\%   & 7.53\%  & 14.82\% & \multirow{4}{*}{$\mathcal{R}_{\mathcal{N}}$} & \multirow{4}{*}{11.86\%}  \\ 
\cline{2-7}
                    & MNSE                        & 51.25\%  & 11.72\%  & 11.68\%  & 10.40\% & 15.65\% &                                              &                           \\ 
\cline{1-7}
\multirow{3}{*}{9}  & $\mathcal{B}^2$             & 0.47\%  & \textcolor{tmi_blue_color}{\textbf{5.43}}\%   & 6.52\%   & 9.23\%  & 1.08\%  &                                              &                           \\ 
\cline{2-7}
                    & $\mathcal{R}_{\mathcal{N}}$ & 87.94\% & \textcolor{cadmiumgreen}{\textbf{10.89}}\%  & 18.71\%  & 7.27\%  & 17.43\% &                                              &                           \\ 
\cline{2-9}
                    & MNSE                        & 88.41\%  & 16.32\% & 25.22\%  & 16.50\% & 18.51\% & \multirow{4}{*}{MNSE}                        & \multirow{4}{*}{11.95\%}  \\ 
\cline{1-7}
\multirow{3}{*}{3}  & $\mathcal{B}^2$             & 1.43\%   & 18.73\%  & 10.58\%  & \textcolor{tmi_blue_color}{\textbf{7.95}}\%  & 2.27\%  &                                              &                           \\ 
\cline{2-7}
                    & $\mathcal{R}_{\mathcal{N}}$ & 284.05\% & 8.66\%  & 26.20\%  & \textcolor{cadmiumgreen}{\textbf{14.29}}\% & 47.98\% &                                              &                           \\ 
\cline{2-7}
                    & MNSE                        & 285.48\% & 27.39\%  & 36.77\%  & 22.25\% & 50.25\% &                                              &                           \\
\bottomrule
\end{tabular}
\end{table*}

In general, there is a trade-off between bias and residual noise in denoising methods. When a clean image is the target, methods seek to reduce the RN as much as possible whereas the bias increases. On the other hand, in dose restoration, which is the case of this work, it is desirable to achieve the same RN as the FD with the low bias as possible. To evaluate both these metrics separately, we used the decomposition on the MNSE, as done before in~\cite{borges2018restoration, shan2021lossfunction}, where RN is calculated as follow:

\begin{equation}
    \mathcal{R}_{\mathcal{N}} = \frac{1}{m\times n}\sum_{i}^{m\times n}{\frac{\mathbb{V}\left(z^{'}_{i}\right)}{y_{i}}}
\end{equation}

\noindent and bias:

\begin{equation}
    \mathcal{B}^2 = \left[ \frac{1}{m\times n}\sum_{i}^{m\times n}{\frac{\left(\mathbb{E}\{z^{'}_{i}\} - y_{i}\right)^2}{y_{i}}} \right] - \frac{\mathcal{R}_{\mathcal{N}}}{p}
\end{equation}

\noindent where $m$ is the number of rows, $n$ is the number of columns, $\mathbb{V}$ indicates the pixel-wise variance along the set of $z^{'}$ image realizations, $\mathbb{E}$ indicates the pixel-wise expectation, $y$ is the noise-free image and $p$ is the number of realizations in the $z^{'}$ set.

We evaluated the MNSE in the anthropomorphic breast phantom. To generate the pseudo-ground-truth, we used 10 images within the FD ones. For each dose reduction factor, 10 images were used and also 10 for the FD. The metric was only evaluated inside the breast phantom and we adjuted the mean value of all images and restorations using an affine transformation between the input image and the pseudo-ground-truth.

As the networks are based on the convolution operation, and it is known that this operation might induce noise correlations in the final image. We measured the power spectrum (PS)~\cite{wu2012spectral, kavuri2020relative} to investigate the impact of this correlation in the restored image, in the following form: 

\begin{equation}
    PS(u,v) = \frac{\Delta x \Delta y}{n_r \, s_{rx} \, s_{ry}} \sum_{k}^{n_r}\left | \mathcal{F}\mathcal{F}\mathcal{T}\{ h(x,y) \times u_k(x,y)\} \right |^2
\end{equation}

\noindent where $\Delta x$ and $\Delta y$ are the pixel size in mm the $x$ and $y$ direction, respectively, $n_r$ is the number of extracted ROIs, $s_{rx}$ and $s_{ry}$ are the ROI number of pixels in the $x$ and $y$ direction, respectively, $h$ is a Hanning window and $\mathcal{F}\mathcal{F}\mathcal{T}$ indicates the discrete Fourier transform. We normalized the PS using the large are signal, as following: 

\begin{equation}
    NPS(u,v) = \frac{PS}{LAS^2}
\end{equation}

\noindent where $LAS$ is calculated as the mean pixel value inside the segmented breast. Finally, the 1D plot was calculated as a radially mean of the 2D spectrum. 

\section{Results \& Discussions}
\label{sec:Results_Discussions}

\begin{table*}[ht]
\centering
\caption{MNSE analysis with its decomposition on bias and RN for the HBL approach and comparison with LD, FD and MB results. Values highlighted in blue mean the lowest bias for all restorations. In green, RN values which are closest to the value from FD images.}
\label{tab:MNSEdecomHBL}
\begin{tabular}{clcccccccc} 
\toprule
mAs                 & Metric                      & Noisy    & DL-N2S  & DL-VST-0.01  & DL-VST-0.34 & DL-VST-0.95 & MB      & \multicolumn{2}{l}{Full-dose}                                            \\ 
\hline
\multirow{3}{*}{30} & $\mathcal{B}^2$             & 0.15\%   & 0.34\%  & \textcolor{tmi_blue_color}{\textbf{0.24}}\%  & 0.30\%      & 0.39\%      & 0.26\%  & \multirow{4}{*}{$\mathcal{B}^2$}             & \multirow{4}{*}{0.09\%}   \\ 
\cline{2-8}
                    & $\mathcal{R}_{\mathcal{N}}$ & 24.86\%  & 14.23\% & 15.73\% & 13.77\%      & \textcolor{cadmiumgreen}{\textbf{11.78}}\%      & 13.23\% &                                              &                           \\ 
\cline{2-8}
                    & MNSE                        & 25.02\%  & 14.57\% & 15.98\% & 14.07\%      & 12.16\%      & 13.50\% &                                              &                           \\ 
\cline{1-8}
\multirow{3}{*}{15} & $\mathcal{B}^2$             & 0.45\%   & 1.08\%  & \textcolor{tmi_blue_color}{\textbf{0.82}}\%       & 0.98\%      & 1.23\%      & \textcolor{tmi_blue_color}{\textbf{0.82}}\%  &                                              &                           \\ 
\cline{2-10}
                    & $\mathcal{R}_{\mathcal{N}}$ & 50.80\%  & 16.83\% & 21.92\%       & 17.07\%      & \textcolor{cadmiumgreen}{\textbf{12.76}}\%      & 14.82\% & \multirow{4}{*}{$\mathcal{R}_{\mathcal{N}}$} & \multirow{4}{*}{11.86\%}  \\ 
\cline{2-8}
                    & MNSE                        & 51.25\%  & 17.92\% & 22.74\%       & 18.05\%      & 13.98\%      & 15.65\% &                                              &                           \\ 
\cline{1-8}
\multirow{3}{*}{9}  & $\mathcal{B}^2$             & 0.47\%   & 1.29\%  & \textcolor{tmi_blue_color}{\textbf{1.04}}\%       & 1.27\%      & 1.64\%      & 1.08\%  &                                              &                           \\ 
\cline{2-8}
                    & $\mathcal{R}_{\mathcal{N}}$ & 87.94\%  & 21.57\% & 30.01\%       & 22.21\%      & \textcolor{cadmiumgreen}{\textbf{15.99}}\%      & 17.43\% &                                              &                           \\ 
\cline{2-10}
                    & MNSE                        & 88.41\%  & 22.86\% & 31.05\%       & 23.48\%      & 17.63\%     & 18.51\% & \multirow{4}{*}{MNSE}                        & \multirow{4}{*}{11.95\%}  \\ 
\cline{1-8}
\multirow{3}{*}{3}  & $\mathcal{B}^2$             & 1.43\%   & 2.48\%  & \textcolor{tmi_blue_color}{\textbf{2.21}}\%       & 2.56\%      & 3.11\%      & 2.27\%  &                                              &                           \\ 
\cline{2-8}
                    & $\mathcal{R}_{\mathcal{N}}$ & 284.05\% & \textcolor{cadmiumgreen}{\textbf{41.36}}\% & 65.43\%       & 48.23\%      & 42.30\%      & 47.98\% &                                              &                           \\ 
\cline{2-8}
                    & MNSE                        & 285.48\% & 43.84\% & 67.65\%       & 50.79\%      & 45.40\%      & 50.25\% &                                              &                           \\
\bottomrule
\end{tabular}
\end{table*}

In this section, we present the visual comparisons and the quantitative results regarding both DBL  and HBL approaches. We show the results as an ablation study, \ie, first DBL is presented such that no priors are added and the network only learns from data. In sequence, the HBL is presented, where we introduce the N2S network representing the training only in the VST, \ie, the forward VST, its inverse and the weighted sum were not taken into account in the loss function. Finally, we present the VST network within the $\lambda$ tuning, illustrating improvements in both visual and quantitative results. 

All the results are regarding the test set, where these cases represent a real scenario, \ie, the LD images were obtained directly in the DBT system. 

\subsection{DBL}

\subsubsection{Visual Comparison}

Figure~\ref{fig:DBL_roi_proj_phantom} illustrates the restoration of LD projections by different networks in distinct dose reductions. From top to bottom, each row represents the reduction factor at 50\%, 25\%, 15\% and 5\%, respectively. The first to last column represents, respectively, the LD projections, FFDM, FFDM with transfer-learning, DBT, MB and the FD images. 

FFDM shows an interesting case, where the network was only trained on images particularly for this system and even though performed well when applied in the DBT test set. This indicates that the network generalizes well even for different systems. Although DBT and FFDM modalities use the same equipment, they have key differences such as detector size, radiation dose per image and anti-scatter grid that affect image formation. In the two lowest dose cases, the network performance is limited and the image quality is very degraded when we compare with the MB approach.   

\begin{figure}[h]
    \centering
	\subfloat[]{\includegraphics[scale=0.171]{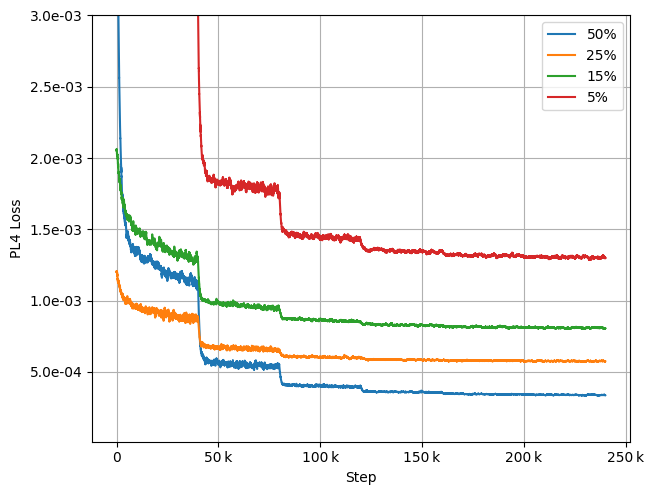}}
	\subfloat[]{\includegraphics[scale=0.171]{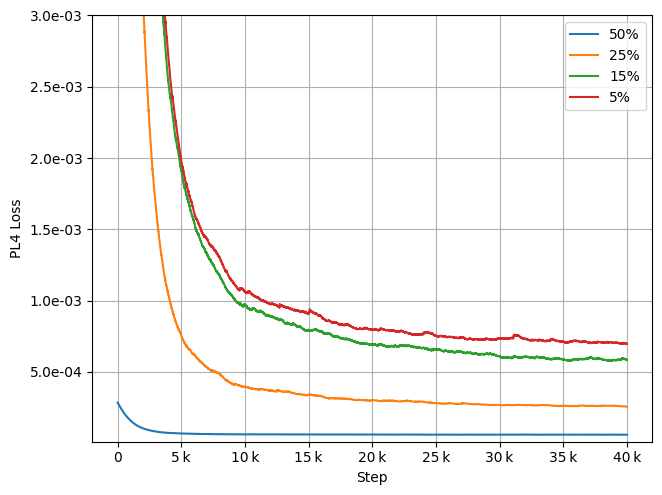}}
	\subfloat[]{\includegraphics[scale=0.171]{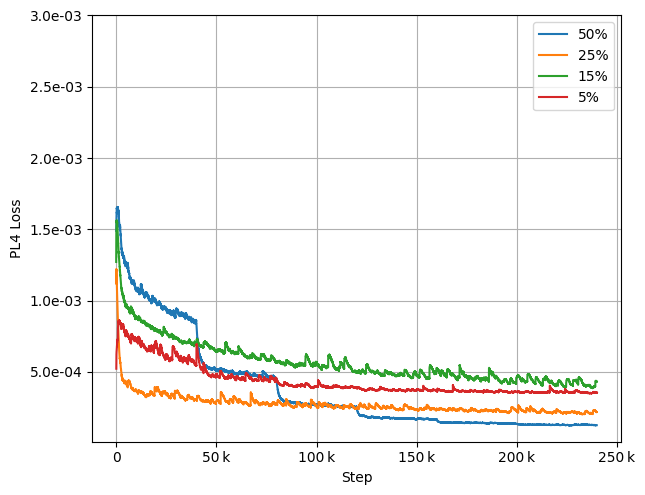}}
	\caption{Illustration of the loss function for the (a) FFDM network, (b) FFDM with transfer-learning and (c) DBT network. We used the exponential moving average with the weight equals 0.999 to better show the losses.}
	\label{fig:lossFunctions}
\end{figure}

In visual terms, the FFDM* network did not benefit from the transfer-learning technique. No real improvements can be seen for neither dose reductions, although its loss function decreased until DBT loss levels, as Fig.~\ref{fig:lossFunctions} illustrates. 

DBT achieved better results for 25\% when compared with FFDM cases, however, in the 15\% case, the underlying signal was degraded, as we can see in the small bright point in the image center. 

In all 50\% cases, the restorations achieved similar results as the MB and the FD. This was expected and showed in our previous work~\cite{shan2021lossfunction}. For other reduction factors, the network did not perform well. We claim that for these lower doses, PL4 does not preserve image details as the input becomes noisier, although it reaches convergence as Fig.~\ref{fig:lossFunctions} shows. One option might be adding an MSE or MAE loss in addition to the PL, so it enforces image fidelity. The drawback of this procedure is that it would be more time-consuming since PL already takes a long training time. Another option would be training the networks after image reconstruction, as the resulting image would be a combination of 15 projections, for the Hologic systems, yielding an image with less noise. 

\subsubsection{Quantitative Analysis}

Table~\ref{tab:MNSEdecomDBL} provides an objective analysis, in terms of MNSE and its decomposition on bias and RN, for LD and FD acquisitions, DL restorations and also the MB method. We highlighted in blue the lowest value of bias within DL restorations and in green the closest RN to the FD value. 

The network trained only on FFDM data had better results for both bias and RN for 50\% and 15\% reduction factors. This could be partially explained by the fact that FFDM images have a higher signal-to-noise ratio (SNR) in comparison with individual DBT projections, so even though the FFDM network was trained on LD images, \eg, 50\%, these LD images contain an SNR higher than the individuals 50\% DBT projections. This makes the learning process easier, so the network does not have to perform aggressive denoising to achieve FD image properties.

DBT network achieved the lowest bias in the 25\% and 5\% cases. However, it only yield the closest RN in the lowest dose. FFDM with with transfer learning only achieved good results in the RN for the 25\% reduction factor. 

In general, MB performed better compared with DL methods in DBL mode. We also can see that in this mode, there is no control over how the network behaves regarding denoising properties. 

\subsection{HBL}

\begin{figure}[h]
    \centering
	\includegraphics[width=1\linewidth]{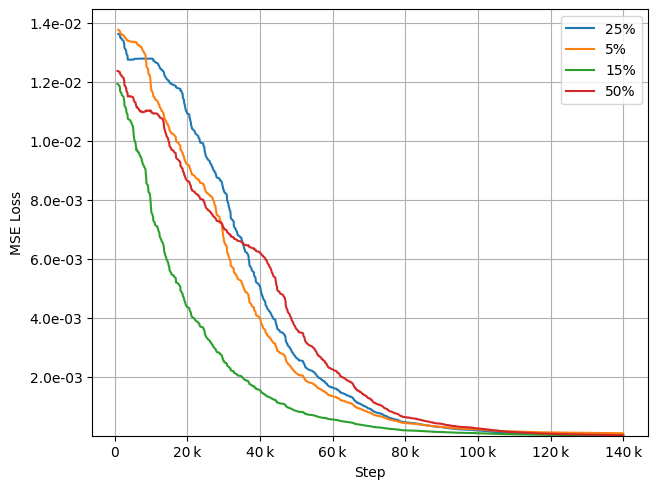}
	\caption{Illustration of the MSE loss function for the N2S network. We used the exponential moving average with the weight equals 0.992 to better show the losses.}
	\label{fig:lossFunctionN2S}
\end{figure}

Although the MSE loss function for the N2S network reached the same values for all reduction factors, as illustrated by Fig.~\ref{fig:lossFunctionN2S}, we selected the network that was trained with 50\% dose reduction, as this network yield less bias. 

\subsubsection{$\lambda_{RN}$ Optimization}

As previously mentioned, there is a trade-off in image restoration when denoising an image between bias and RN. We optimized the $\lambda_{RN}$ hyper-parameter to evaluate the values where the network seeks low bias, RN matching, or an intermediate point. 

\begin{figure}
    \centering
    \includegraphics[width=1\linewidth]{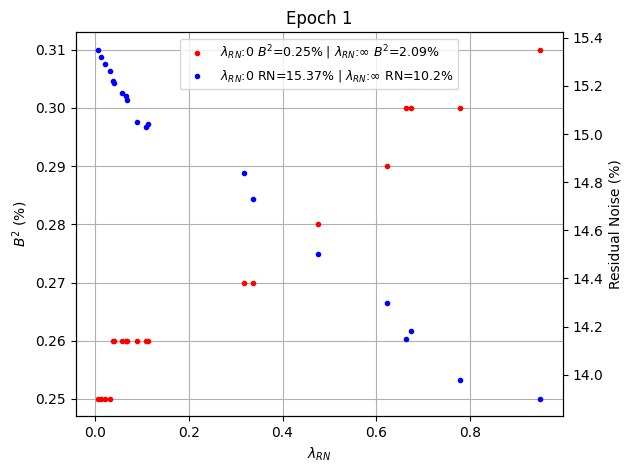}
    \caption{$\lambda_{RN}$ tuning plot on the first epoch. The graphic illustrates the network operation points in terms of bias and RN for each $\lambda_{RN}$ value.}
    \label{fig:wRNxb2xRN_ep01}
\end{figure}

\begin{figure}
    \centering
    \includegraphics[width=1\linewidth]{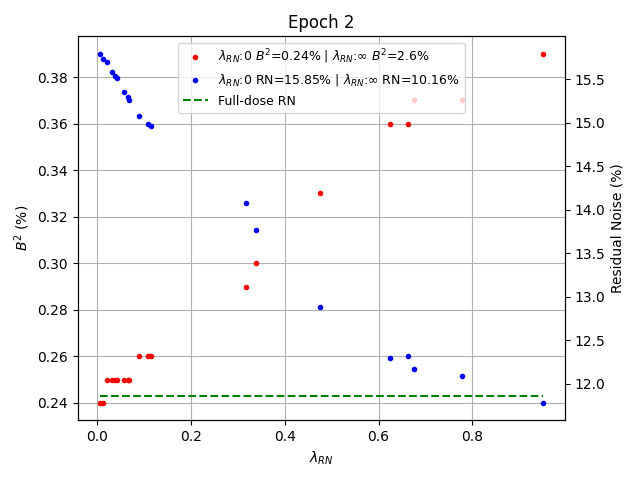}
    \caption{$\lambda_{RN}$ tuning plot on the second epoch. The graphic illustrates the network operation points in terms of bias and RN for each $\lambda_{RN}$ value..}
    \label{fig:wRNxb2xRN_ep02}
\end{figure}

Figures~\ref{fig:wRNxb2xRN_ep01} and~\ref{fig:wRNxb2xRN_ep02} show the results of $\mathcal{B}^2$ and $\mathcal{R}_{\mathcal{N}}$ for different $\lambda_{RN}$ values, when training for 1 and 2 epochs, respectively. In general, we can observe, objectively, the trade-off between the measured metrics and also that the $\lambda_{RN}$ regarding the intermediate point decreased in the second epoch. This point represents the hyper-parameter value where the network achieves the lowest bias and RN. Approximately, for epoch 1, $\lambda_{RN} = 0.42$ and $\lambda_{RN} = 0.38$ for epoch 2. It can also be inferred that when RN is preferred, the network has strong denoising properties because the loss forces a noise variance math between input and target images. On the other hand, and the loss is bias seek, the network prioritizes the underlying signal, yielding cautious denoising and higher noise variance. Theoretically, the lowest bias that could be achieved is when the output image is the same as the input. In terms of RN, the lowest it can be achieved would be performing total denoising seeking the underlying signal. In DBT image restoration, we seek a match between RN of FD and LD images, with low bias as possible. 

We also measured two different network operation points, where the loss function only contains the bias term, \ie, $\lambda_{RN} = 0$, and the other point where it only contains the RN term, \ie, $\lambda_{RN} = \infty$.     

\subsubsection{Visual Comparison}

Figure~\ref{fig:HBL_roi_proj_phantom} illustrates the results regarding the HBL approach. The same ROI from Fig.~\ref{fig:DBL_roi_proj_phantom} is shown, now for N2S and VST DL methods. For the VST network, we choose three $\lambda_{RN}$ values (0.01, 0.34 and 0.95) to show different points of operation. These points were considered because they represent the case with the lowest bias, intermediate point and closest RN. In general, all DL methods achieve satisfactory results which are favorable or comparable with MB ones. 

\begin{figure*}[ht]
	\centering	
	\subfloat[50\%]{\includegraphics[scale=0.27]{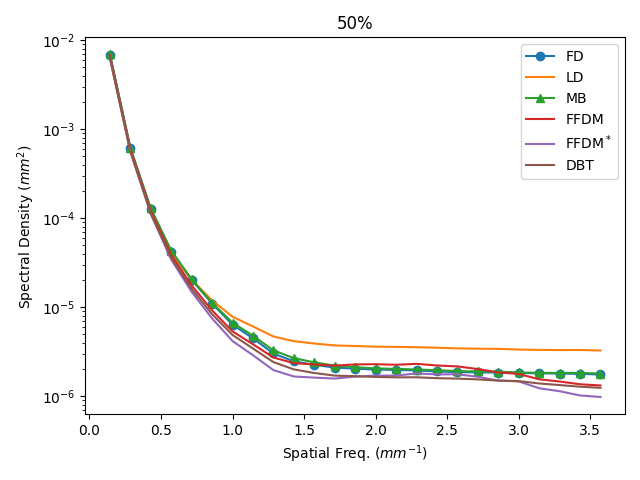}}
	\subfloat[25\%]{\includegraphics[scale=0.27]{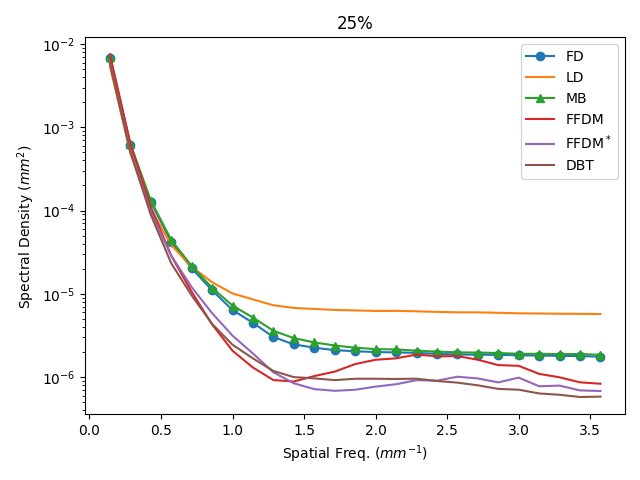}}
	\subfloat[15\%]{\includegraphics[scale=0.27]{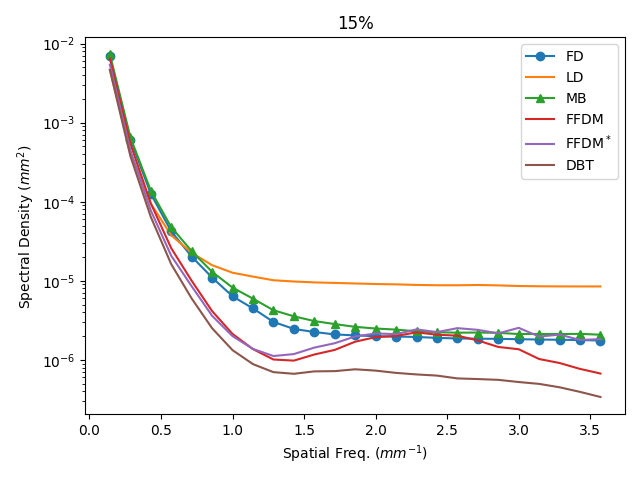}}
	\subfloat[05\%]{\includegraphics[scale=0.27]{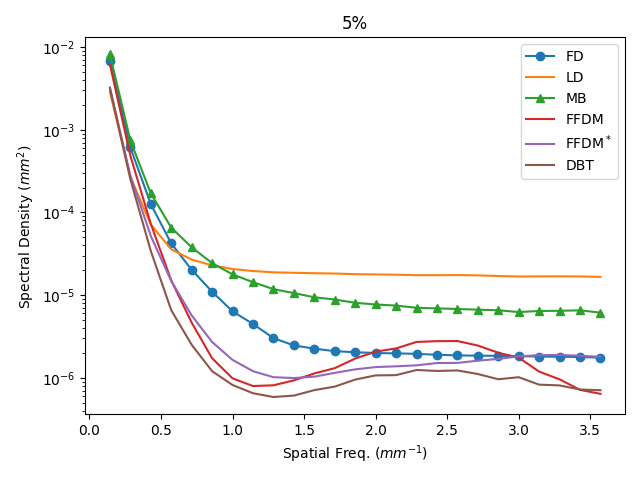}}
	\caption{Graphics illustrating the NPS for all the restorations in DBL mode, respective to the reduction factors of (a) 50\%, (b) 25\%, (c), 15\% and (d) 5\%.}
    \label{fig:PS-DBL}
\end{figure*}

\begin{figure*}[ht]
	\centering	
	\subfloat[50\%]{\includegraphics[scale=0.27]{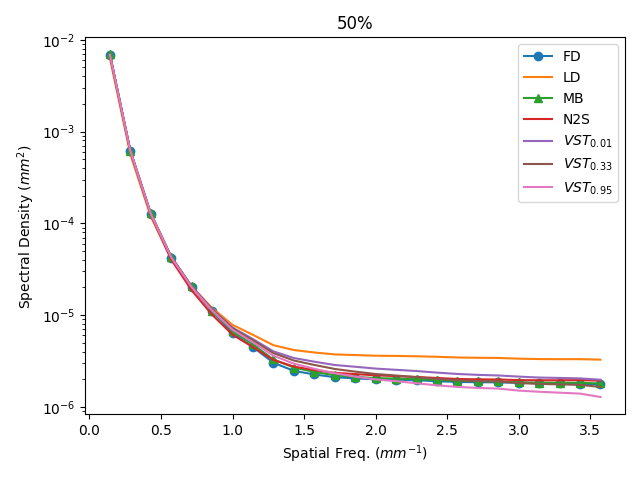}}
	\subfloat[25\%]{\includegraphics[scale=0.27]{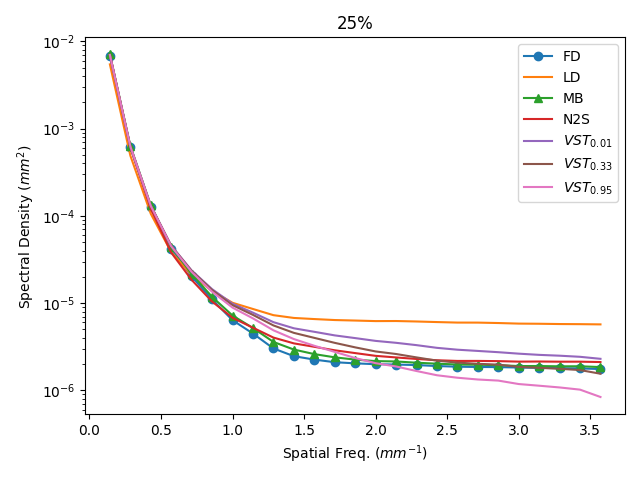}}
	\subfloat[15\%]{\includegraphics[scale=0.27]{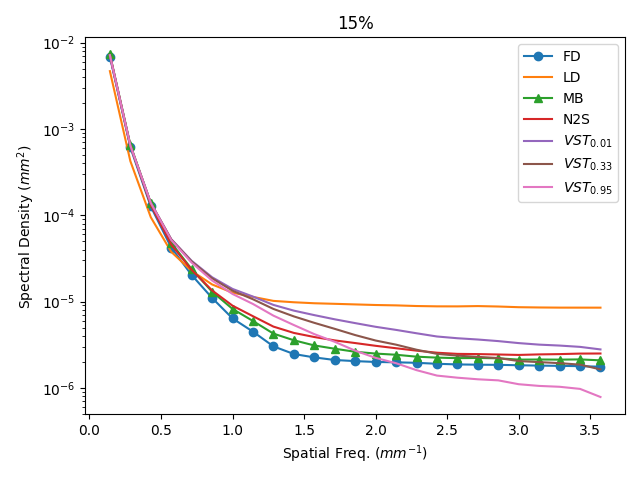}}
	\subfloat[05\%]{\includegraphics[scale=0.27]{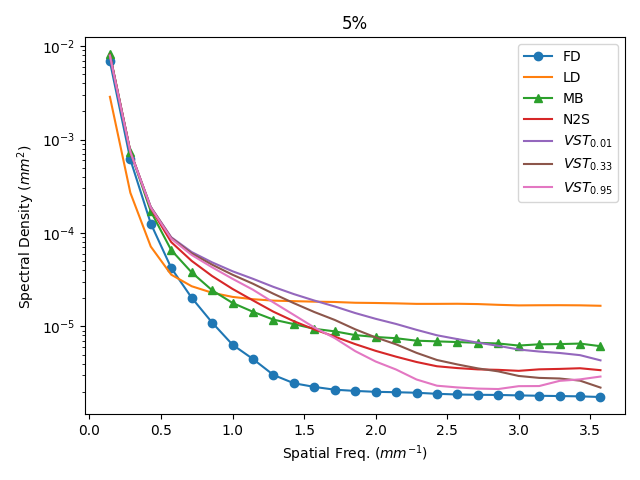}}
	\caption{Graphics illustrating the NPS for all the restorations in HBL mode, respective to the reduction factors of (a) 50\%, (b) 25\%, (c), 15\% and (d) 5\%.}
    \label{fig:PS-HBL}
\end{figure*}

N2S by itself benefited from training in the VST domain, where noise has approximately unit variance and already achieved good results. Although we used this framework for training the network, it is possible to use any other architecture and framework for the denoising job. The limitation of this method is that it is not possible to control the network operation point in terms of bias and RN. 

In the VST network, we can observe that the $\lambda_{RN}$ directly affects image results. For the highest value, it is possible to see that the network has strong denoising properties. Differently, in the lowest value, we can see much more noise when compared to the previous case. In terms of bias, \ie, signal preservation, we can notice that the tinny bright spot is sharper in VST-0.01 oppositely to the VST-0.95 where it is more blurred. VST-0.34 shows a network case that compromises in terms of preservation of the underlying signal and RN matching. 

\subsubsection{Quantitative Analysis}

Table~\ref{tab:MNSEdecomHBL} contains the MNSE, bias and RN results concerning the HBL mode. Once more, we highlighted the values with the lowest bias and closest RN for all reduction factors. Different from Tab.~\ref{tab:MNSEdecomDBL}, now we considered the MB method. 

With respect to bias, the VST-0.01 network achieved the lowest values for all reduction factors, except in the 25\% case, where it achieves the same as the MB. In terms of RN, the VST-0.01 network accomplished RN values closest to the FD one, except for the lowest reduction factor. Both facts demonstrate the effectiveness of the BRN loss function to fine-tune the network. Also, it shows that it is possible to control the network operation concerning denoising strength, accomplishing the specific results, unlike the DBL mode. Moreover, adding prior knowledge to the training process enhanced the restoration quantitative results, as shown by N2S by itself.      

Another advantage of HBL mode is the training time, as illustrated by Tab.~\ref{tab:time}. For DBL, using the PL4, an average of 86.5h were spent for training purposes, among 7.5h for pre-training with L1 and 79 for training with PL4. PL4* illustrates the time for the transfer learning technique. However, this network was previously trained with FFDM data, also spending around 86.5h. HBL only takes approximately 8h for training in the N2S framework, as it uses the low-time cost MSE loss function. Finally, the fine-tunning process with BRN loss only takes 3min for an epoch. It is also worth noting that PL4 requires the VGG network to be loaded in GPU memory, demanding more memory and, consequently decreasing image batch size. 

\begin{table}[h]
\centering
\caption{Training time, in hours, for both DBL and HBL modes. L1 represents the pre-training process in DBL, PL4 the training process in HBL, PL$4^*$ the fine-tuning for FFDM, N2S the training in the VST domain and BRN the fine-tuning for the VST network.}
\label{tab:time}
\begin{tabular}{lc}
\toprule
Model & Time (h)  \\
\hline
L1    &  7.5      \\
PL4   &  79      \\
PL4*  &  13      \\
N2S   &  8.5      \\
BRN   &  0.1     \\
\bottomrule
\end{tabular}
\end{table}

As observed in Fig.~\ref{fig:DBL_roi_proj_phantom}, some DL restorations impose correlations on the resulted image. To objectively measure it, we evaluated the PS in all restorations, as illustrated by Fig.~\ref{fig:PS-DBL} and~\ref{fig:PS-HBL} for DBL and HBL, respectively.  

We can see that DBL methods diverge from LD and FD curves, especially for high-frequency (HF) values. Even for low frequency, the curves start to move away in 25\% and lower doses. Also, in HF, the curve does not have a flat behavior as LD and FD do, indicating that the network correlates the noise. 

Different from DBL mode, HBL approximates well from FD and LD curves, both in low and HF values. N2S network had the best performance, due to its training in the VST domain, where noise has approximately unity variance. VST networks introduce some correlation as the fine-tuning was performed in the image domain, where noise does not have a well-defined behavior in terms of variance. However, this correlation does have wave tendency as observed in DBL HF values. It is also worth noting that the MB method achieved the best performance in terms of PS for all frequencies. 

\section{Conclusion}
\label{sec:Conclusion}

In conclusion, CNNs designed with prior knowledge on its architecture benefit from it and have superior results from those without it. For the VST network, we achieved comparable or superior results from the MB method in terms of bias and RN. When HBL was evaluated against DBL mode, it achieved superior results in training time, bias, RN, PS and also requires less GPU memory. 

This work has some limitations. We did not evaluate DBL with different losses and combinations of them. This would require tuning weight parameters between these losses. Also, we restricted the network architecture for the modified ResNet presented in our previous work. Further evaluation with different architectures such as U-net~\cite{ronneberger2015u}, original ResNet~\cite{he2016deep}, WGAN~\cite{arjovsky2017wasserstein}, Transformers~\cite{dosovitskiy2020image}, etc., is fundamental. Also, we could evaluate different frameworks for training the network in the VST domain. Another limitation respects the equipment parameters estimation necessity for the GAT and weighted sum layers. Future works could investigate performing the GAT, its inverse and the wSum as trainable layers and compare the results. Also, there is a necessity to inspect the $\lambda_{RN}$ parameter in terms of lesion diagnosis with radiologists. This would answer the question of whether is a low bias image with a slightly high RN is preferred or an RN match between restored image and FD, with a loss of bias.

Finally, we claim that the presented framework for training networks for image restoration can be expanded for other areas in the medical image field and also for different areas such as natural images and video processing.

\section*{Acknowledgment}

This work was supported in part by the National Council for Scientific and Technological Development (CNPq), by the S\~{a}o Paulo Research Foundation (FAPESP grant 2021/12673-6 and 2018/19888-5), by the \textit{Coordenação de Aperfeiçoamento de Pessoal de Nível Superior} (CAPES finance code 001), and National Institute of Health (R01EB026646, R01CA233888, R01CA237267, and R01HL151561).

The authors would like to thank the Institute of Radiology (InRad), in particular Dra. Denise Yanikian Nersissian and Éric Francisco Scolastici, for providing the clinical data. The authors also would like to thank Dr. Andrew D. A. Maidment, from the University of Pennsylvania, for making the anthropomorphic breast phantom images available for this work. 


\bibliographystyle{unsrt}

\begin{thebibliography}{10}

\bibitem{lai2020digital}
Yi-Chen Lai, Kimberly~M Ray, James~G Mainprize, Tatiana Kelil, and Bonnie~N
  Joe.
\newblock Digital breast tomosynthesis: technique and common artifacts.
\newblock {\em Journal of Breast Imaging}, 2(6):615--628, 2020.

\bibitem{das2010penalized}
Mini Das, Howard~C Gifford, J~Michael O'Connor, and Stephen~J Glick.
\newblock Penalized maximum likelihood reconstruction for improved
  microcalcification detection in breast tomosynthesis.
\newblock {\em IEEE Transactions on Medical Imaging}, 30(4):904--914, 2010.

\bibitem{xu2015statistical}
Shiyu Xu, Jianping Lu, Otto Zhou, and Ying Chen.
\newblock Statistical iterative reconstruction to improve image quality for
  digital breast tomosynthesis.
\newblock {\em Medical physics}, 42(9):5377--5390, 2015.

\bibitem{zheng2017detector}
Jiabei Zheng, Jeffrey~A Fessler, and Heang-Ping Chan.
\newblock Detector blur and correlated noise modeling for digital breast
  tomosynthesis reconstruction.
\newblock {\em IEEE transactions on medical imaging}, 37(1):116--127, 2017.

\bibitem{wu2012dose}
Gang Wu, James~G Mainprize, and Martin~J Yaffe.
\newblock Dose reduction for digital breast tomosynthesis by patch-based
  denoising in reconstruction.
\newblock In {\em International Workshop on Digital Mammography}, pages
  721--728. Springer, 2012.

\bibitem{borges2017pipeline}
Lucas~R Borges, Predrag~R Bakic, Alessandro Foi, Andrew~DA Maidment, and
  Marcelo~AC Vieira.
\newblock Pipeline for effective denoising of digital mammography and digital
  breast tomosynthesis.
\newblock In {\em Medical Imaging 2017: Physics of Medical Imaging}, volume
  10132, page 1013206. International Society for Optics and Photonics, 2017.

\bibitem{borges2018restoration}
Lucas~R Borges, Lucio Azzari, Predrag~R Bakic, Andrew~DA Maidment, Marcelo~AC
  Vieira, and Alessandro Foi.
\newblock Restoration of low-dose digital breast tomosynthesis.
\newblock {\em Measurement Science and Technology}, 29(6):064003, 2018.

\bibitem{wu2017iterative}
Dufan Wu, Kyungsang Kim, Georges El~Fakhri, and Quanzheng Li.
\newblock Iterative low-dose {CT} reconstruction with priors trained by
  artificial neural network.
\newblock {\em IEEE Transactions on Medical Imaging}, 36(12):2479--2486, 2017.

\bibitem{kang2017deep}
Eunhee Kang, Junhong Min, and Jong~Chul Ye.
\newblock A deep convolutional neural network using directional wavelets for
  low-dose x-ray {CT} reconstruction.
\newblock {\em Medical Physics}, 44(10):e360--e375, 2017.

\bibitem{chen2017low1}
Hu~Chen, Yi~Zhang, Mannudeep~K Kalra, Feng Lin, Yang Chen, Peixi Liao, Jiliu
  Zhou, and Ge~Wang.
\newblock Low-dose {CT} with a residual encoder-decoder convolutional neural
  network.
\newblock {\em IEEE Transactions on Medical Imaging}, 36(12):2524--2535, 2017.

\bibitem{wolterink2017generative}
Jelmer~M Wolterink, Tim Leiner, Max~A Viergever, and Ivana I{\v{s}}gum.
\newblock Generative adversarial networks for noise reduction in low-dose {CT}.
\newblock {\em IEEE Transactions on Medical Imaging}, 36(12):2536--2545, 2017.

\bibitem{chen2017low2}
Hu~Chen, Yi~Zhang, Weihua Zhang, Peixi Liao, Ke~Li, Jiliu Zhou, and Ge~Wang.
\newblock Low-dose {CT} via convolutional neural network.
\newblock {\em Biomedical Optics Express}, 8(2):679--694, 2017.

\bibitem{kang2018deep}
Eunhee Kang, Won Chang, Jaejun Yoo, and Jong~Chul Ye.
\newblock Deep convolutional framelet denosing for low-dose {CT} via wavelet
  residual network.
\newblock {\em IEEE Transactions on Medical Imaging}, 37(6):1358--1369, 2018.

\bibitem{shan20183}
Hongming Shan, Yi~Zhang, Qingsong Yang, Uwe Kruger, Mannudeep~K Kalra, Ling
  Sun, Wenxiang Cong, and Ge~Wang.
\newblock {3-D} convolutional encoder-decoder network for low-dose {CT} via
  transfer learning from a {2-D} trained network.
\newblock {\em IEEE Transactions on Medical Imaging}, 37(6):1522--1534, 2018.

\bibitem{shan2019competitive}
Hongming Shan, Atul Padole, Fatemeh Homayounieh, Uwe Kruger, Ruhani~Doda Khera,
  Chayanin Nitiwarangkul, Mannudeep~K Kalra, and Ge~Wang.
\newblock Competitive performance of a modularized deep neural network compared
  to commercial algorithms for low-dose {CT} image reconstruction.
\newblock {\em Nature Machine Intelligence}, 1(6):269--276, 2019.

\bibitem{yin2019domain}
Xiangrui Yin, Qianlong Zhao, Jin Liu, Wei Yang, Jian Yang, Guotao Quan, Yang
  Chen, Huazhong Shu, Limin Luo, and Jean-Louis Coatrieux.
\newblock Domain progressive {3D} residual convolution network to improve
  low-dose {CT} imaging.
\newblock {\em IEEE Transactions on Medical Imaging}, 38(12):2903--2913, 2019.

\bibitem{wu2021drone}
Weiwen Wu, Dianlin Hu, Chuang Niu, Hengyong Yu, Varut Vardhanabhuti, and
  Ge~Wang.
\newblock Drone: Dual-domain residual-based optimization network for
  sparse-view ct reconstruction.
\newblock {\em IEEE Transactions on Medical Imaging}, 2021.

\bibitem{gao2021deep}
Mingjie Gao, Jeffrey~A Fessler, and Heang-Ping Chan.
\newblock Deep convolutional neural network with adversarial training for
  denoising digital breast tomosynthesis images.
\newblock {\em IEEE Transactions on Medical Imaging}, 2021.

\bibitem{liu2018radiation}
Junchi Liu, Amin Zarshenas, Syed~Ammar Qadir, Limin Yang, Laurie Fajardo, and
  Kenji Suzuki.
\newblock Radiation dose reduction in digital breast tomosynthesis ({DBT}) by
  means of neural network convolution ({NNC}) deep learning.
\newblock In {\em 14th international workshop on breast imaging (IWBI 2018)},
  volume 10718, page 1071814. International Society for Optics and Photonics,
  2018.

\bibitem{sahu2019using}
Pranjal Sahu, Hailiang Huang, Wei Zhao, and Hong Qin.
\newblock Using virtual digital breast tomosynthesis for de-noising of low-dose
  projection images.
\newblock In {\em 2019 IEEE 16th International Symposium on Biomedical Imaging
  (ISBI 2019)}, pages 1647--1651. IEEE, 2019.

\bibitem{chen2018learn}
Hu~Chen, Yi~Zhang, Yunjin Chen, Junfeng Zhang, Weihua Zhang, Huaiqiang Sun,
  Yang Lv, Peixi Liao, Jiliu Zhou, and Ge~Wang.
\newblock Learn: Learned experts’ assessment-based reconstruction network for
  sparse-data {CT}.
\newblock {\em IEEE Transactions on Medical Imaging}, 37(6):1333--1347, 2018.

\bibitem{wurfl2018deep}
Tobias W{\"u}rfl, Mathis Hoffmann, Vincent Christlein, Katharina Breininger,
  Yixin Huang, Mathias Unberath, and Andreas~K Maier.
\newblock Deep learning computed tomography: Learning projection-domain weights
  from image domain in limited angle problems.
\newblock {\em IEEE Transactions on Medical Imaging}, 37(6):1454--1463, 2018.

\bibitem{adler2018learned}
Jonas Adler and Ozan {\"O}ktem.
\newblock Learned primal-dual reconstruction.
\newblock {\em IEEE Transactions on Medical Imaging}, 37(6):1322--1332, 2018.

\bibitem{gong2018pet}
Kuang Gong, Ciprian Catana, Jinyi Qi, and Quanzheng Li.
\newblock {PET} image reconstruction using deep image prior.
\newblock {\em IEEE Transactions on Medical Imaging}, 38(7):1655--1665, 2018.

\bibitem{zhang2019vst}
Minghui Zhang, Fengqin Zhang, Qiegen Liu, and Shanshan Wang.
\newblock Vst-net: variance-stabilizing transformation inspired network for
  poisson denoising.
\newblock {\em Journal of Visual Communication and Image Representation},
  62:12--22, 2019.

\bibitem{niu2020noise2sim}
Chuang Niu and Ge~Wang.
\newblock Noise2sim--similarity-based self-learning for image denoising.
\newblock {\em arXiv preprint arXiv:2011.03384}, 2020.

\bibitem{anscombe1948transformation}
Francis~J Anscombe.
\newblock The transformation of poisson, binomial and negative-binomial data.
\newblock {\em Biometrika}, 35(3/4):246--254, 1948.

\bibitem{nagare2021bias}
Madhuri Nagare, Roman Melnyk, Obaidullah Rahman, Ken~D Sauer, and Charles~A
  Bouman.
\newblock A bias-reducing loss function for ct image denoising.
\newblock In {\em ICASSP 2021-2021 IEEE International Conference on Acoustics,
  Speech and Signal Processing (ICASSP)}, pages 1175--1179. IEEE, 2021.

\bibitem{starck1998image}
Jean-Luc Starck, Fionn~D Murtagh, and Albert Bijaoui.
\newblock {\em Image processing and data analysis: the multiscale approach}.
\newblock Cambridge University Press, 1998.

\bibitem{makitalo2012optimal}
Markku Makitalo and Alessandro Foi.
\newblock Optimal inversion of the generalized anscombe transformation for
  poisson-gaussian noise.
\newblock {\em IEEE transactions on image processing}, 22(1):91--103, 2012.

\bibitem{shan2021lossfunction}
Hongming Shan, Rodrigo~B. Vimieiro, Lucas~R. Borges, Marcelo A.~C. Vieira, and
  Ge~Wang.
\newblock Impact of loss functions on the performance of a deep neural network
  designed to restore low-dose digital mammography.
\newblock {\em arXiv preprint arXiv:}, 2021.

\bibitem{zhao2016loss}
Hang Zhao, Orazio Gallo, Iuri Frosio, and Jan Kautz.
\newblock Loss functions for image restoration with neural networks.
\newblock {\em IEEE Transactions on Computational Imaging}, 3(1):47--57, 2016.

\bibitem{borges2016method}
Lucas~R Borges, Helder CR~de Oliveira, Polyana~F Nunes, Predrag~R Bakic,
  Andrew~DA Maidment, and Marcelo~AC Vieira.
\newblock Method for simulating dose reduction in digital mammography using the
  {Anscombe} transformation.
\newblock {\em Medical Physics}, 43(6Part1):2704--2714, 2016.

\bibitem{borges2017method}
Lucas~R Borges, Igor Guerrero, Predrag~R Bakic, Alessandro Foi, Andrew~DA
  Maidment, and Marcelo~AC Vieira.
\newblock Method for simulating dose reduction in digital breast tomosynthesis.
\newblock {\em IEEE Transactions on Medical Imaging}, 36(11):2331--2342, 2017.

\bibitem{BakicVCT}
Predrag~R Bakic, David Higginbotham, Bruno Barufaldi, and Andrew D.~A.
  Maidment.
\newblock {The open-source virtual clinical trial project}.
\newblock Sourceforge \url{ https://sourceforge.net/projects/openvct}.
\newblock (Accessed: 04 Fev 2022).

\bibitem{borges2019noise}
Lucas~R Borges, Bruno Barufaldi, Renato~F Caron, Predrag~R Bakic, Alessandro
  Foi, Andrew~DA Maidment, and Marcelo~AC Vieira.
\newblock Noise models for virtual clinical trials of digital breast
  tomosynthesis.
\newblock {\em Medical physics}, 46(6):2683--2689, 2019.

\bibitem{carton2011development}
Ann-Katherine Carton, Predrag Bakic, Christer Ullberg, Helen Derand, and
  Andrew~DA Maidment.
\newblock Development of a physical {3D} anthropomorphic breast phantom.
\newblock {\em Medical Physics}, 38(2):891--896, 2011.

\bibitem{dabov2007image}
Kostadin Dabov, Alessandro Foi, Vladimir Katkovnik, and Karen Egiazarian.
\newblock Image denoising by sparse {3-D} transform-domain collaborative
  filtering.
\newblock {\em IEEE Transactions on Image Processing}, 16(8):2080--2095, 2007.

\bibitem{wu2012spectral}
Gang Wu, James~G Mainprize, and Martin~J Yaffe.
\newblock Spectral analysis of mammographic images using a multitaper method.
\newblock {\em Medical physics}, 39(2):801--810, 2012.

\bibitem{kavuri2020relative}
Amar Kavuri and Mini Das.
\newblock Relative contributions of anatomical and quantum noise in signal
  detection and perception of tomographic digital breast images.
\newblock {\em IEEE transactions on medical imaging}, 39(11):3321--3330, 2020.

\bibitem{ronneberger2015u}
Olaf Ronneberger, Philipp Fischer, and Thomas Brox.
\newblock {U-net: Convolutional} networks for biomedical image segmentation.
\newblock In {\em International Conference on Medical image computing and
  computer-assisted intervention}, pages 234--241. Springer, 2015.

\bibitem{he2016deep}
Kaiming He, Xiangyu Zhang, Shaoqing Ren, and Jian Sun.
\newblock Deep residual learning for image recognition.
\newblock In {\em Proceedings of the IEEE conference on computer vision and
  pattern recognition}, pages 770--778, 2016.

\bibitem{arjovsky2017wasserstein}
Martin Arjovsky, Soumith Chintala, and L{\'e}on Bottou.
\newblock {Wasserstein} generative adversarial networks.
\newblock In {\em International conference on machine learning}, pages
  214--223. PMLR, 2017.

\bibitem{dosovitskiy2020image}
Alexey Dosovitskiy, Lucas Beyer, Alexander Kolesnikov, Dirk Weissenborn,
  Xiaohua Zhai, Thomas Unterthiner, Mostafa Dehghani, Matthias Minderer, Georg
  Heigold, Sylvain Gelly, et~al.
\newblock An image is worth 16x16 words: Transformers for image recognition at
  scale.
\newblock {\em arXiv preprint arXiv:2010.11929}, 2020.

\end{thebibliography}

\end{document}